\documentclass{aa}  

\usepackage{graphicx}
\usepackage{txfonts}

\begin{document}

   \title{Potential contributions of Pop III and intermediate-mass Pop II stars to cosmic chemical enrichment}

   \subtitle{}

   \author{Lia C. Corazza
          \inst{1},
          Oswaldo D. Miranda
          \inst{1}
          \and
          Carlos A. Wuensche
          \inst{1}}

   \institute{Instituto Nacional de Pesquisas Espaciais,
              Av. dos Astronautas 1758 -- Jardim da Granja \\
                S\~ao Jos\'e dos Campos, SP 12227--010, Brazil
             }

   \date{Received ; accepted }

 
  \abstract
   {We propose a semi-analytic model that is developed to understand the cosmological evolution of the mean metallicity in the Universe. In particular, we study the contributions of Population III (Pop III) and Population II (Pop II) stars to the production of $\mathrm{Fe,~Si,~Zn, ~Ni,~P, ~Mg, ~Al, ~S, ~C, ~N}$, and $\mathrm{~O}$.}
   {We aim to quantify the roles of two different models in the chemical enrichment of the Universe. The first model (A) considers both stars with Pop III and Pop II yields. For the second model (B), the yields involved are only for Pop II stars.}
   {We start by describing the cosmic star formation rate (CSFR) through an adaptation of a scenario developed within the hierarchical scenario of structure formation with a Press-Schechter-like formalism. We adapt the formalism to implement the CSFR to the standard chemical evolution scenario to investigate the course of chemical evolution on a cosmological basis. Calculations start at redshift $z\sim 20$, and we compare the results of our two models with data from damped Lyman-$\alpha$ systems (DLAs), and globular clusters (GCs).}
   {Our main results find that metal production in the Universe occurred very early, quickly increasing with the formation of the first stars. When comparing results for [Fe/H] with observations from GCs, yields of Pop II stars are not enough to explain the observed chemical abundances, requiring stars with physical properties similar those expected from Pop III stars.}
   {Our semi-analytic model can deliver consistent results for the evolution of cosmic metallicities. Our results show that the chemical enrichment in the early Universe is rapid, and at redshift $\sim 12.5$, the metallicity reaches $10^{-4}\, Z_{\sun}$ for the model that includes Pop III stars. In addition, we explore values for the initial mass function (IMF) within the range $[0.85, 1.85]$.}


\titlerunning{Pop III and intermediate-mass Pop II
stars to cosmic chemical enrichment}

\authorrunning{Corazza, Miranda and Wuensche}

\maketitle
%

\section{Introduction}
\label{sec:intro}

In order to understand the chemical evolution of the entire Universe, it is vital that we understand the global mechanisms that have dominated the production of chemical elements since the Big Bang. Primordial nucleosynthesis is responsible for the synthesis of deuterium, $^{3}$He, $^{4}$He, and traces of $^{7}$Li, and it ceased after the first few minutes of the existence of the Universe. After that period, a global chemical enrichment would resume only when stellar nucleosynthesis started inside the nuclei of stars.

Two classes of stars are of primary interest for the chemical evolution scenario: primordial, metal-free stars, known as first stars or Population III (Pop III) stars, and second-generation, more enriched stars, known as Population II (Pop II) stars. It is believed that Pop III stars had very unusual proprieties, and despite intense observational efforts, these stars have not yet been observed, although a few candidates have been proposed (e.g.,
\citealp{2012ApJ...761...85K,2015ApJ...808..139S,2020MNRAS.494L..81V}). Thus, researchers have been combining efforts to build consistent modeling of these stars in recent decades \citep{heger2002nucleosynthetic,schaerer2002,chieffilimongi,hegerwoosley2010,Takahashi}. Results indicate that they would have very large masses, between $100$ and $200\,\mathrm{M}_{\sun}$, or even higher, around $500$ to $1000\, \mathrm{M}_{\sun}$ (\citealp{Ohkubo2006}), mainly due to the lack of metals in the gas, making cooling processes very inefficient \citep{DawnChems,2013Hirano}.

Also, the question about their initial mass function (IMF) and their role in the reionization of the Universe \citep{2000ApJ...540...39A,2001ApJ...548...19N,1997ApJ...486..581G,2000MNRAS.315L..51C,2000ApJ...528L..65T,2001MNRAS.324..381C,BarkanaLoeb,Venkatesan,Tumlinson2004,2006Sci...313..931B}, among other properties, have been discussed to better understand the physics of this high-mass stellar population. 

In terms of chemical production, the above results indicate they were substantially important. For instance, Pop III stars with masses from $140$ to $260\, \textrm{M}_{\sun}$ produced huge amounts of metals. They ended their lives as pair-instability supernovae (PISNe), injecting highly enriched material back into the interstellar medium (ISM) and intergalactic medium (IGM), and leaving no remnants behind after a complete disruptive process \citep{heger2002nucleosynthetic,Takahashi}. 

After the first generation of Pop III stars started to die, injecting large amounts of metals into the ISM and IGM, Pop II stars started to form. With more and more enriched material available, cooling processes started to become more efficient, giving origin to less massive stars with physical proprieties close to those observed today, and also extensively modeled mainly according to their masses and metallicities \citep{chieffilimongi,2005IAUS..228..315K,campbelllattanzio2008,karakas2010,doherty2013,doherty2014}.  

Cosmological chemical evolution models have also been largely explored. There are several different models that seek to evaluate different aspects related to the chemical enrichment, such as the evolution of the mass-metallicity relation \citep{2016MNRAS.456.2140M,2019MNRAS.484.5587T}, the establishment of a critical value for the metallicity of the Universe, enabling the transition from Pop III to Pop II stars \citep{2001MNRAS.328..969B,2003BL,Fang2004,MatteucciEarly,SantoroShull2006,TornatoreTransition,MaioTransition,Schneider2010}, Hypernovae (HNe) feedback \citep{2007MNRAS.376.1465K}, the role of galactic outflows \citep{Dave2007}, the chemical properties of local galaxies based on their formation through the hierarchical model of structure formation \citep{calura2009}, the evolution of N abundance in the Universe and the reason for a large dispersion in observational data \citep{VangioniNitrogen}, and the influence of dark matter (DM) halos on the gas reservoir available for star formation \citep{2013ApJ...772..119L}, among other examples of interesting contributions to the study of the Universe through its chemical enrichment. 

The models can be generically classified into semi-analytic and hydrodynamic simulations. Nevertheless, there is increasing uncertainty connected to the chemical evolution  as we move from local to cosmological scales, which are independent of the analytic or computational modeling. From the small scale represented by nuclear reaction rates and stellar masses to larger, galactic, and cosmological scales, there is a cumulative uncertainty since each scale carries its own set of considerations and uncertainties. For a specific discussion on this subject, see, in particular, \citet{cote2016}. These models help discuss general and particular aspects of metallicity evolution in cosmological terms, but several do not detail the contributions to the evolution of single elements. Moreover, comparing observations with high-redshift simulations is a challenge. Based upon the points mentioned above, we propose a semi-analytic model to investigate the contributions of Pop III and Pop II stars to the cosmological evolution of single elements across the redshift interval $0 \le z \lesssim 20$, not including the details of hydrodynamic simulations, and taking into account different perspectives to compare our results with observations in a range of different redshifts.\\

We start in Sect. \ref{sec2}, introducing and justifying the choices for the cosmological background, which is going to be the basis for the chemical evolution model: we describe the model developed by \citet{pereiraemiranda}, and the incorporated changes in the scenario that allow for an adequate coupling of the star formation model with the equations of the chemical evolution of the Universe. We address the adapted cosmological model as Corazza, Miranda \& Wuensche (hereinafter represented as CMW along with the text). In addition, the modifications of the model introduced in this work allow for a better adjustment of the cosmic star formation rate (CSFR) to the observational data available up to redshift $\sim 10$, and satisfy all the points studied by \citet{Gribel2017} in their unified model connecting the CSFR with the local star formation.

We also adapted the chemical models developed over the past 40 years for the Galaxy (see, e.g., \citealp{tinsleylarson1978,tinsleylarson1980,Matteucci2016}). This adaptation allows us to build an adequate model for the chemical enrichment of the Universe. Implementing chemical yields for stars with masses between $0.85$ and $260\, \textrm{M}_{\sun}$ and metallicities from $0$ up to $Z=0.02\;Z_{\sun}$ allows us to provide, in Sect. \ref{results}, several results, data comparison, and discussions about the cosmic evolution of 11 chemical elements: iron (Fe), silicon (Si), zinc (Zn), nickel (Ni), phosphorus (P), magnesium (Mg), aluminum (Al), sulfur (S), carbon (C), nitrogen (N), and oxygen (O). We draw the conclusions in Sect. \ref{final}.

\section{Methodology}\label{sec2}

In this section, we show how we obtain the CSFR from the process of large-scale structure formation and how we build a scenario that describes the cosmic chemical enrichment from redshift $\sim 20$ to the present. The first DM halos decoupled from the Universe's expansion, collapsing and virializing, probably between the end of recombination and redshift $\sim 20$. The potential wells of these first halos generated the conditions for the baryonic matter to flow into these structures, agglomerate, and form the first stars. The characterization of the cosmological star formation and the consequent chemical enrichment of the Universe is, in this way, connected to the DM halo formation within a given mass range and as function of the redshift.

\subsection{Cosmological scenario}
\label{subsec:csfr}

Dark matter halos drag the baryonic matter into their interiors. We can describe this process through the adaptation of the formalism developed originally by \citet{PS74}, which allows us to directly estimate the fraction of baryons ($f_\mathrm{b}$) incorporated into the halos:

\begin{equation}\label{fraction}
f_\mathrm{b} (z) = \frac{\int_{M_\mathrm{min}}^{M_\mathrm{max}}\,f(M,z)\,M\,dM}{\int_{0}^{\infty} f(M,z)\,M\,dM},
\end{equation}
where
\begin{equation}
df(M,z) = \frac{\rho_\mathrm{m}}{M}\frac{d\ln{\sigma^{-1}}}{dM}f_\mathrm{ST}(\sigma)\, dM
\end{equation}
is the number of DM halos per comoving volume at a given redshift within the mass interval $[M, M + dM]$, and $\rho_\mathrm{m}$ is the matter density of the Universe.

The halo mass function, $f_\mathrm{ST}(\sigma)$, proposed by \citet{shethtormen}, is:
\begin{equation}
    f_\mathrm{ST}(\sigma) = 0.3222 \sqrt{\frac{2a}{\pi}}\frac{\delta_\mathrm{c}}{\sigma} \exp \left( \frac{-a\delta_\mathrm{c}^{2}}{2\sigma^{2}}\right) \left[ 1 + \left( \frac{\sigma^{2}}{a\delta_\mathrm{c}^{2}} \right) ^{p}\right],\label{eqn_f}
\end{equation}
with $\delta_\mathrm{c} = 1.686$, $a = 0.707$, $p = 0.3$, and $\sigma(M,z)$ is the variance of the linear density field.

The fact that stars form only in suitably dense structures is parameterized in Eq. (\ref{fraction}) by the threshold mass $M_\mathrm{min}$. We consider $M_\mathrm{min}=10^{6}\,\mathrm{M}_{\sun}$ to be the minimum mass for the first star-forming halos to appear in hierarchical models. The upper limit $M_\mathrm{max}$ can take values up to $\gtrsim 10^{17}\, \mathrm{M}_{\sun}$. This limit is set according to the mass scale of galaxy superclusters \citep{salv2007, pereiraemiranda}, limiting the scale of the largest structures formed in the present Universe. In any case, the results have shown to be weakly dependent on the upper limit if $M_\mathrm{max} > 10^{17}\, \mathrm{M}_{\sun}$.

The function $\sigma(M,z)$ in Eq. (\ref{eqn_f}) can be determined from the power spectrum $P(k)$ smoothed with a spherical top-hat filter function of radius $R$, which, on average, encloses a mass $M$ $(R=[3M/4\pi\rho(z)]^{1/3})$. Thus,

\begin{equation}
\sigma^{2}(M,z) = \frac{D^{2}(z)}{2\pi^{2}} \int_{0}^{\infty}{k^{2}\, P(k)\, W^{2}(k,M)\,dk},
\end{equation}
where $W(k,M)$ is the top-hat filter in the $k$-space:

\begin{equation}
W(k,M) = \frac{3}{(kR)^{3}}\,[\sin(kR)-k R\cos(kR)].
\end{equation}

The dependence on redshift comes from the growth factor $D(z)$, that is, $\sigma(M,z)=\sigma(M,0)D(z)$. Here we use the analytical approach for $D(z)$ as derived by \citet{1992Carrol}.

The rate at which fluctuations grow on different scales depends on the interplay between self-gravitation, pressure support, and damping processes. All of these processes are part of the power spectrum given by $P(k) \propto k^{n_\mathrm{p}}$ (see, e.g., \citealp{Gribel2017} for details).

From these equations, it is possible to determine how halos of different masses decouple from the Universe's expansion and how baryonic matter is gradually incorporated into the center of the virialized halos. Structures more massive than $\sim 10^{6}\, \mathrm{M}_{\sun}$ are formed at later times, as the redshift decreases. Thus, as more halos are formed, more baryonic matter flows into these structures, generating conditions for star formation. This allows us to define the baryon accretion rate as:

\begin{equation}\label{baryon}
 a_\mathrm{b}(t) = \Omega_{0,\mathrm{b}}\,\rho_{0,\mathrm{c}} \left(\frac{dt}{dz}\right)^{-1} \left| \frac{df_\mathrm{b}}{dz} \right|,
\end{equation}
where $\Omega_{0,\mathrm{b}}$ is the baryonic density parameter at $z=0$, $\rho_{0, \mathrm{c}} = 3H_{0}^{2}/8\pi G $ is the critical density of the Universe ($H_{0} = 100\, h\, \mathrm{km}\,\mathrm{s}^{-1}\,\mathrm{Mpc}^{-1}$ is the value of the Hubble parameter at the current time), and

\begin{equation}
    \frac{dt}{dz} = \frac{1}{H_{0}\, (1+z)\sqrt{\Omega_{0,\Lambda}+\Omega_{0,\mathrm{m}}\,(1+z)^{3}}}.
\end{equation}

The cosmological framework described through the set of equations presented above is similar to the one used by different authors (e.g., \citealp{Daigne_2006}; \citealp{pereiraemiranda}; \citealp{Tan_2016}; \citealp{Gribel2017}; \citealp{VangioniNitrogen}).

\subsection{Cosmic star formation rate}
\label{subsec:csfr_2}

Once we set the cosmological framework, it is possible to compute the CSFR by incorporating the IMF and the star formation rate (SFR). In particular, the number of stars formed per unit of mass ($m$), volume ($V$), and time ($t$) is given by

\begin{equation}
 \frac{d^{3} N(m,V, t)}{dmdVdt} = \varphi(m)\, \psi (t),\label{dn3}
\end{equation}
where $\psi(t)\propto \rho_\mathrm{g}^{\alpha}$ corresponds to the SFR ($\rho_\mathrm{g}$ is the gas density). It should be noted that $\psi(t)$ follows the functional form known as Schmidt's law \citep{Schmidt}. On the other hand, the IMF is given by $\varphi(m)\propto m^{-(1+x)}$ and its functional form with $x=1.35$ is called as Sapeter's IMF \citep{salpeter}.\\

The IMF of the first stars is still an open question. We see that  the Salpeter IMF favors the formation of low-mass stars, and various authors adopted it in their chemical evolution models (see, e.g., \citealp{calura,2006AIPC..847..371C,2012Casey,2016Shu,Fraser2017,VangioniNitrogen}) while some others (see, e.g., \citealp{2001ApJ...548...19N, 2006MNRAS.369..825S, Ma2016}) also allow for the possibility of a top-heavy or bi-modal
IMF.

In our study, we consider $x=1.35$ as the reference value. However, to identify the influence of the IMF exponent on the chemical enrichment of the Universe, we also considered four other values, nominally, $0.85$, $1.0$, $1.7$, and $1.85$ allowing the formation of a 
higher (the first two values) or smaller (the last two values) number of high-mass stars when compared to the reference value $1.35$. We also used $\alpha=1$ in agreement with \citet{Gribel2017}, which shows that different properties from the star formation regions in the Galaxy, including the so-called Larson's law, can be well reproduced with $\alpha=1$.

Therefore, Eq. (\ref{dn3}) describes the number of stars formed within the DM halos that aggregate and concentrate baryons in their centers. A fraction of the mass in stars is ejected (through stellar winds and supernovae, for example) and returned to the ISM formed by these structures. The ejected mass fraction is given by:

\begin{equation}
 \frac{d^{2}M_\mathrm{ej}}{dVdt} = \int^{m_\mathrm{ s}}_{m{(t)}} (m-m_\mathrm{r})\, \psi(t-\tau_\mathrm{m})\,\varphi(m)\,dm,
 \label{eject}
\end{equation}

\noindent where $m(t)$ is the stellar mass whose lifetime is equal to $t$, and $m_\mathrm{r}$ represents the mass of the remnant, which depends both on the progenitor mass ($m$) and on the environment metallicity $(Z)$. The star formation is taken at the time ($t - \tau_\mathrm{m}$), where $\tau_\mathrm{m}$, also a function of the metallicity $(Z)$, is the lifetime of a star of mass $m$.

We used the results of \citet{Spera2015} to obtain the masses of the stellar remnants $(m_\mathrm{r})$ as functions of the metallicity and the initial stellar masses. The authors obtained their results from \textrm{SEVN} (population-synthesis code) coupled with the \textrm{PARSEC} code for stellar evolution tracks. In particular, in this work we use the fitting formulas presented in Appendix C of \citet{Spera2015}.

Concerning the parameter $\tau_\mathrm{m} $, we use the metallicity-dependent formula given by \citet{Raiteri1996}:

\begin{equation}
    \log\,\tau_\mathrm{m} = a_{0}(Z) + a_{1}(Z)\log \left(\frac{M_{\star}}{\mathrm{M}_{\sun}}\right) + a_{2}(Z) \left[\log \left( \frac{M_{\star}}{\mathrm{M}_{\sun}}\right) \right]^{2},\label{lifetime}
\end{equation}
where $\tau_\mathrm{m}$ is expressed in years, and the metallicity-dependent coefficients are (see \citealp{Raiteri1996} for details):

\begin{equation}
a_{0}(Z) = 10.13+0.07547\, \log Z-0.008084\, (\log Z)^{2},
\end{equation}
\begin{equation}
a_{1}(Z) = -4.424-0.7939\, \log Z-0.1187\, (\log Z)^{2} \ ,
\end{equation}
\begin{equation}
a_{2}(Z) = 1.262+0.3385\, \log Z + 0.05417\, (\log Z)^{2},
\end{equation}
and $Z$ is the absolute metallicity.

It is worth stressing that $\tau_\mathrm{m}$ determined by Eq. (\ref{lifetime}) has an excellent agreement when compared to the stellar lifetimes presented in Table 2 of \citet{refId0} for different values of mass and metallicity. In particular, the difference between the results for $\tau_\mathrm{m}$ is lower than $5\%$, which has little effect on our results.

Following this formalism and combining the previous equations, we derive the equation that governs the total gas density $\rho_\mathrm{g}$ in the halos:

\begin{equation}\label{gas_total}
 \dot\rho_\mathrm{g} = -\frac{d^{2}M_{\star}}{dVdt} + \frac{d^{2}M_\mathrm{ej}}{dVdt} + a_\mathrm{b}(t),
\end{equation}
where the term $a_\mathrm{b}(t)$ gives, for the halos, a matter of primordial composition. The system becomes closed without the term $a_\mathrm{b}(t)$ in Eq. (\ref{gas_total}). Thus, this term corresponds to a primordial gas infall in the structures in formation. In other words, it describes the primordial baryonic matter that is captured by the potential wells generated by the halos.

On the other hand, the first term on the right side gives the mass of gas converted to stars per unit of volume and time. By Schmidt's law, we have:

\begin{equation}
\psi(t) = \frac{d^{2}M_{\star}}{dVdt} = k\,\rho_\mathrm{g}.
\end{equation}

\noindent It should be noted that the term $k$ is the inverse of timescale for star formation, that is, $k = 1/\tau_\mathrm{s}$.

The total gas density can be calculated by numerical integration of Eq. (\ref{gas_total}), providing values for $\rho_\mathrm{g}(t)$ at each time $t$ or redshift $z$ as long as the $\tau_\mathrm{s}$ parameter is set. The initial condition is zero gas density at $t = 0$ $(z=20)$ for solving Eq. (\ref{gas_total}). Moreover, there are some steps for obtaining the correct characterization of the function $\rho_\mathrm{g}$. First of all, it is necessary to determine the function $\tau_\mathrm{s}$. This parameter is related to the CSFR via Schmidt's Law, that is, $\dot\rho_{\star}$ is directly proportional to the gas density and inversely proportional to the characteristic timescale for the conversion of gas in stars. Second, if all the gas entering the system, plus the gas returning to the system through Eq. (\ref{eject}), is converted into stars, there is an overabundance of both stars and metals. Thus, $\dot\rho_{\star}$ must also be dependent on a parameter that measures the efficiency ($<\varepsilon_{\star}>$) in which gas is converted into stars.

From the above considerations, we have:

\begin{equation}
   \dot\rho_{\star} (z) = <\varepsilon_{\star}> \frac{\rho_\mathrm{g}}{\tau_\mathrm{s}}.\label{csfr_final} 
\end{equation}

Once the CSFR is determined, the $a_\mathrm{b}$ function is fixed by the structure formation scenario, and the IMF is determined by the choice of the $x$ exponent, and then it will be possible to determine the function $\rho_\mathrm{g}$ by Eq. (\ref{gas_total}).

The aforementioned steps are essential to characterize the total gas density function. The calculation algorithm integrates the differential equation (\ref{gas_total}) through the sixth-order Runge-Kutta method. The differential equations for the various chemical elements and total metallicity of the Universe (Sect. \ref{CE}) are solved by the same method.

\subsubsection{Characterizing the functions $<\varepsilon_{\star}>$ and $\tau_\mathrm{s}$ }

The $<\varepsilon_{\star}>$ and $\tau_\mathrm{s}$ functions work together to produce the CSFR with the best fit to the observational data. In particular, the cold gas used to form stars is given by:

\begin{equation}
    \rho_\mathrm{cold}(z) = <\varepsilon_{\star}>\,{\rho_\mathrm{g}(z)},
    \label{effi}
\end{equation}
\noindent where $<\varepsilon_{\star}>$ acts as efficiency for star formation. In principle, $\rho_\mathrm{cold}$ is composed of the sum of two components: molecular gas ($\rho_\mathrm{H2}$) and atomic gas ($\rho_\mathrm{HI}$). This allows us to rewrite $\tau_\mathrm{s}$ as:

\begin{equation}
    \tau_\mathrm{s} (z) = \frac{\rho_\mathrm{cold}}{\dot\rho_{\star}} = \frac{\rho_\mathrm{H2}}{\dot\rho_{\star}} + \frac{\rho_\mathrm{HI}}{\dot\rho_{\star}} = \tau_\mathrm{depl,H2} + \tau_\mathrm{depl,HI},
    \label{taus}
\end{equation}
\noindent with $\tau_\mathrm{depl,H2}$ and $\tau_\mathrm{depl,HI}$ representing, respectively, the depletion scales for molecular and atomic gases.

Equations (\ref{csfr_final}), (\ref{effi}), and (\ref{taus}) are solved together to characterize the CSFR and to determine the correct dependence of the $<\varepsilon_{\star}>$ and $\tau_\mathrm{s} $ functions on the redshift. The constraints associated with these equations are the following: the best adjustment of the $\dot\rho_{\star}(z)$ curve to the observational data available within the range $[0-10]$ in redshift must be produced; $\dot\rho_{\star}$ should be normalize to return the value $\sim 0.016\,\mathrm{M}_{\sun}\,\mathrm{yr}^{-1}\,\mathrm{Mpc}^{-3}$ at $z = 0$, similar to the one determined by \citet{madau2014}; the CSFR peak should be made at redshift $z = 2$. This value was chosen so that $\dot\rho_{\star}$ obtained here is in accordance with the peak of the CSFR used by \citet{VangioniNitrogen} and the one determined by \citet{madau2014}. Furthermore, $<\varepsilon_{\star}>\,\sim 0.01-0.02$ should be produced at $z=0$. This causes the value $<\varepsilon_{\star}>$ to be on the order of $\varepsilon_\mathrm{ff}$, the so-called SFR per free-fall time, inferred for the star-forming regions of the local Universe (see, e.g., \citealp{km2005,Gribel2017}). Lastly,
$\tau_\mathrm{s} (z = 0)\sim 0.5-2.5\, \mathrm{Gyr}$ should be produced, similar to the value of $\tau_\mathrm{dep}$ as inferred by \citet{Schinnerer2016} for the local Universe ($z=0$).

\begin{figure*}\label{fig:csfr1}
\includegraphics[width=\textwidth]{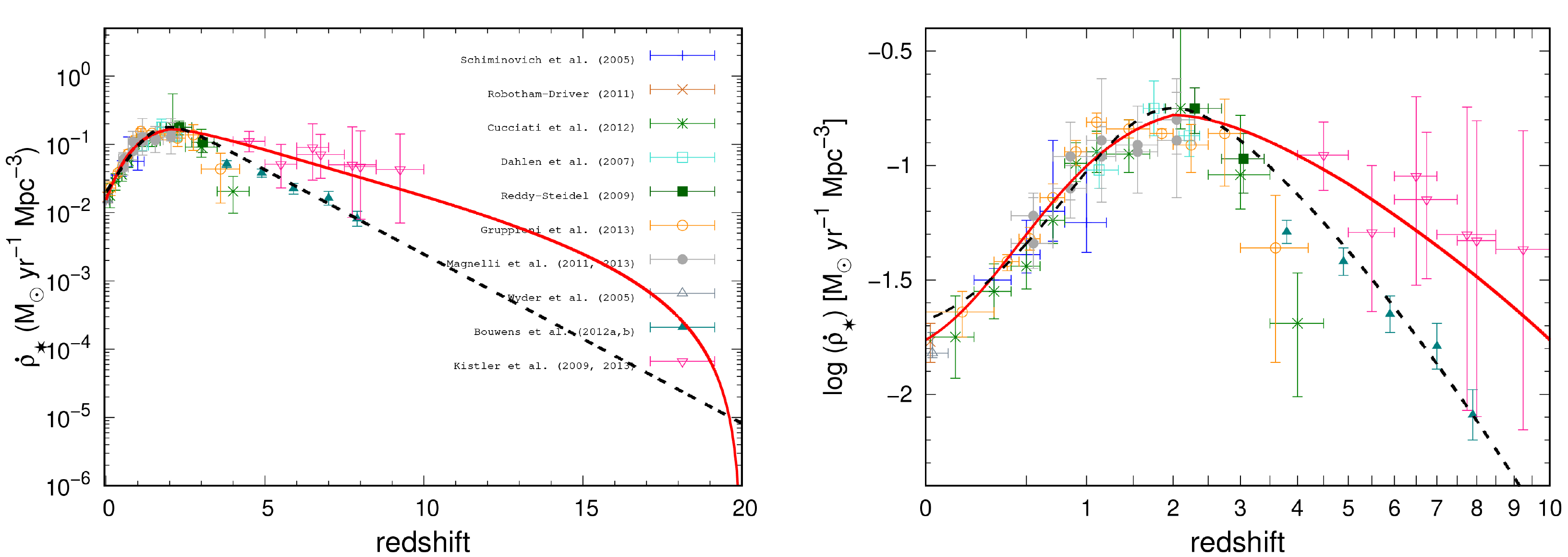}
\caption{Solution for $\dot\rho_{\star}(z)$ as derived in this work (red line) and, using black dashed line, the CSFR used by \citet{VangioniNitrogen} plotted for comparison. Left: Evolution of the CSFR from the local Universe to $z = 20$. Right: Same as for the left panel, but zooming into $0 \le z \le 10$, allowing for a better visualization of the two CSFRs within the range of the available observational data.
Data used in this figure: IR (\citet{mag2011,mag2013} - dark gray filled circles; \citet{gru2013} - dark orange open circles); UV ( \citet{wyder2005} - slate gray open triangle; \citet{schi2005} - blue crosses; \citet{dahlen2007} - turquoise open squares; \citet{rs2009} - dark green filled squares; \citet{rodriver2011} - chocolate cross; \citet{cucci2012} - green stars; \citet{bowens2012b,bowens2012a} - teal filled triangles); and GRB (\citet{kistler2009} - deep pink open triangles).}

\label{fig:csfr}
\end{figure*}

Figure \ref{fig:csfr} shows the CSFR as a function of redshift and its behavior concerning the observational data. For comparison, we also present the CSFR used by \citet{VangioniNitrogen}. Both our CSFR and the one used by \citet{VangioniNitrogen} reach a maximum value at $z = 2$, as mentioned above. Up to redshift $\sim 4$, the two CSFRs have a very similar behavior. We fit both CSFRs to results from IR, UV, and GRB observations, as described in the figure caption.

\begin{figure*}
\centering
\includegraphics[width=\textwidth]{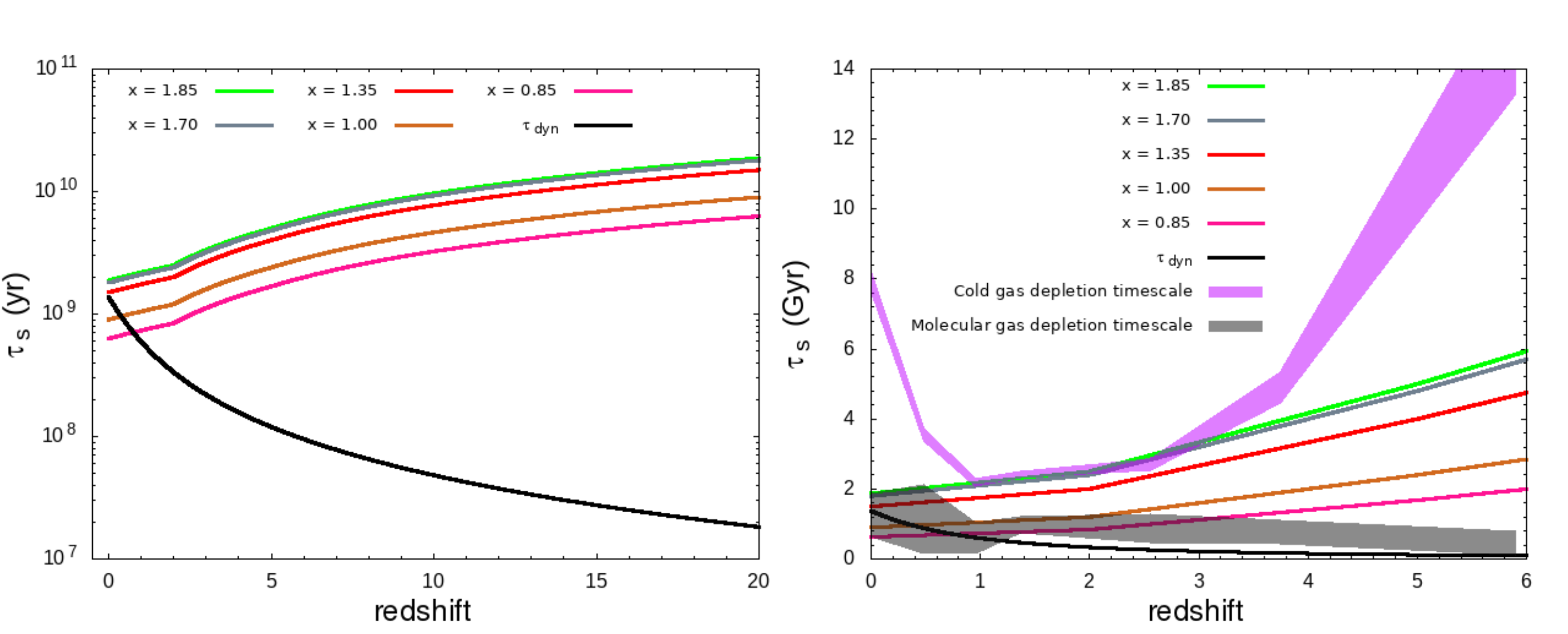}
\caption{
Evolution of the characteristic timescale for star formation as a function of the redshift.  Left: $\tau_\mathrm{s}$ for different IMF exponent and the evolution of the dynamical time ($t_\mathrm{dyn}$), taken to be 10\% of Hubble time. Right: Our results compared with the timescales for the conversion of gas (filled areas) obtained by \citet{peroux} and \citet{maio2022} for molecular and cold (HI+H2) gas.
}
 \label{fig:efi_tau}
\end{figure*}

In Fig. \ref{fig:efi_tau} we present the behavior of the $\tau_\mathrm{s}$ parameter for different values of the $x$ exponent.  Additionally, as determined by \citet{peroux} and \citet{maio2022}, the corresponding $\mathrm{H}_{2}$ depletion time closely follows the dynamical time ($\tau_\mathrm{dyn}$), taken to be 10\% of Hubble time. The curves for $\tau_\mathrm{s}$ and $\tau_\mathrm{dyn} \sim \tau_\mathrm{depl,H2}$ are shown in the left panel. It should be noted that the values for $\tau_\mathrm{s}$ are above $\tau_\mathrm{dyn}$ for higher redshifts, which means that the largest contribution to the characteristic timescale for star formation, in Eq. (18), is due to the atomic gas instead of molecular gas.

At lower redshifts ($z<5$), the curve $\tau_\mathrm{dyn}$ gradually approaches $\tau_\mathrm{s}$, implying a greater contribution of molecular gas to star formation. We note that for the two IMFs we have $\tau_\mathrm{dyn} > \tau_\mathrm{s}$, which happens at redshift $\sim 0.75$ ($\sim 0.38$) to $x=0.85$ ($1.0$). In these cases, the approximation $\tau_\mathrm{dyn} \simeq t_\mathrm{hubble}/10$ is not adequate to describe the characteristic timescale for star formation. As a result, $\tau_\mathrm{depl,H2}$ should be constant. For the other IMFs, $\tau_\mathrm{s}$ is dominated by the molecular depletion time for $z<0.5$.

The right panel of Fig. \ref{fig:efi_tau} shows our results compared with the timescales for the conversion of gas (filled areas) obtained by \citet{peroux} and \citet{maio2022} for molecular and cold ($\mathrm{HI+H2}$) gas. The behavior of $\tau_\mathrm{s}$ between the two filled areas shows that, in our scenario, star formation is fueled by cold gas in atomic form for $z>6$. When the curves describing $\tau_\mathrm{s}$ approach the gray filled area, the contribution of molecular gas becomes gradually dominant to supply star formation.

As commented above, for IMF exponents 0.85 and 1.0, the approximation $\tau_\mathrm{dyn}\simeq t_\mathrm{hubble}/10$ fails for $z<1$ and $\tau_\mathrm{dyn}$ should be constant in order to supply the star formation with gas in molecular form. Our model shows that for $x$ in the range $1.35-1.85$, about 70\% to 90\% of the star formation would come from molecular gas, for $z\lesssim 1$, if $\tau_\mathrm{depl,H2} = \tau_\mathrm{dyn} \simeq t_\mathrm{hubble}/10$.

\begin{figure*}
\centering
\includegraphics[width=\textwidth]{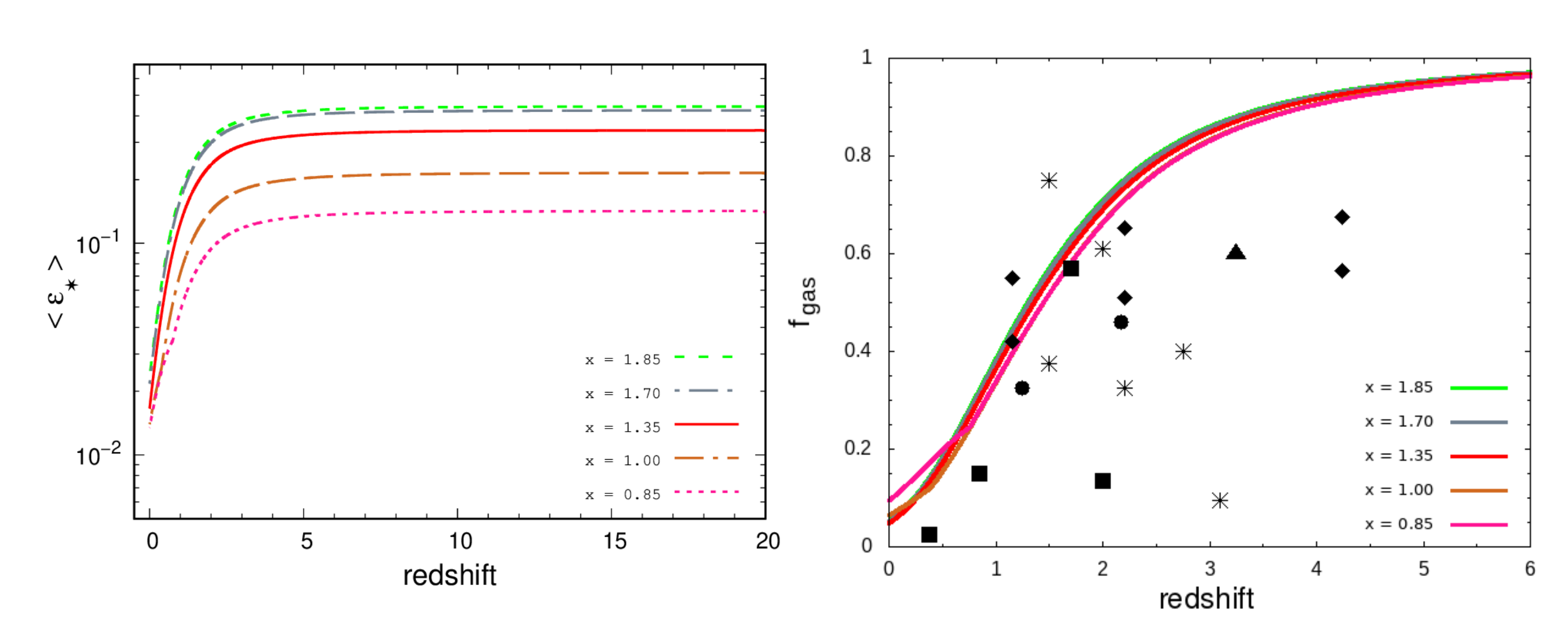}
\caption{
Evolution of the fraction of gas converted into stars for different values of the IMF used in this work. Left: Evolution of the average star formation efficiency (defined as $<\varepsilon_{\star}>$) as a function of $z$. Right: Evolution of the fraction of total cold gas associated with star formation ($f_\mathrm{gas}$). The symbols show the measurements compiled as follows: circles by \citet{Tacconi_2018}, squares by \citet{Scoville_2014}, diamonds by  \citet{Scoville_2016}, stars by \citet{refdz}, and triangles by \citet{Schinnerer2016}. See also Fig. 13b of \citet{hodge2020}.}
 \label{fig:efi_tau_2}
\end{figure*}

The left panel of Fig. \ref{fig:efi_tau_2} shows the evolution of star formation efficiency ($<\varepsilon_{\star}>$) with redshift. This parameter reaches values between $\sim 0.15-0.40$, depending on the value of $x$ for redshifts $z> 3$. Efficiency gradually decreases with redshift reaching values close to $0.01-0.02$ at $z=0$. In the right panel, we present the fraction of total cold gas mass ($f_\mathrm{gas}$) determined by the relation:

\begin{equation}
    f_\mathrm{gas} = \frac{M_\mathrm{gas}}{M_\mathrm{gas}+M_{\star}},
    \label{f_gas}
\end{equation}
where $M_\mathrm{gas}$ is the cold gas mass and $M_{\star}$ is the mass in stars.

Equation (\ref{f_gas}) permits a direct comparison with results obtained by other authors. In particular, \citet{hodge2020} present in their fig. 13b measurements compiled for the cold gas fraction ($\mathrm{H2+HI}$) and compared with scaling relations derived in three different works (\citealp{Scoville2017,Tacconi_2018,Liu_2019}). Our curves for the cold gas fraction show good agreement with the scaling relation $M_\mathrm{gas} \sim M_{\star}^{0.65}$ derived by \citet{Tacconi_2018}.

In Table \ref{initialp}, we summarize the parameters used to obtain the CSFR -- $\dot\rho_{\star}(z)$. It depends on the cosmological parameters $\Omega_{0,\mathrm{m}}$, $\Omega_{0,\mathrm{b}}$, and $\Omega_{0,\Lambda}$, and the parameters related to the formation of large-scale structures of the Universe ($\sigma_{8}$, $n_\mathrm{p}$,  and $M_\mathrm{min}$).

\begin{table}[!ht]
\begin{center}
\caption{Cosmological and structure formation parameters used to obtain the CSFR}
\small
\centering
\begin{tabular}{ccccccccc}
\hline
\hline 
$\Omega_{0,\mathrm{m}}$ & $\Omega_{0,\mathrm{b}}$ & $\Omega_{0,\Lambda}$ & $h$ & $z_\mathrm{i}$ & $\sigma_{8}$ & $n_\mathrm{p}$ & $M_\mathrm{min}(\mathrm{M}_{\sun})$ \\ 
\hline 
0.279& 0.0463 & 0.721 & 0.7 & 20 & 0.84 & 0.967 & $10^{6}$  \\ 
\hline
\end{tabular} 
\label{initialp}
\end{center}
\small \textbf{Note.} $\Omega_{0,\mathrm{m}}$ corresponds to the total matter (baryonic plus DM) density parameter; $\Omega_{0,\mathrm{b}}$ is the baryonic density parameter; $\Omega_{0,\Lambda}$ is the density parameter associated with dark energy (cosmological constant); $h$ is the Hubble constant written as $H_{0}=100\,h\,\mathrm{km}\,\mathrm{s}^{-1}\,\mathrm{Mpc}^{-1}$; $z_\mathrm{i}$ is the redshift at which star formation begins; $\sigma_{8}$ is the normalization of the power spectrum, in other words $\sigma(M,0)$; $n_\mathrm{p}$ is the spectral index of the power spectrum; $M_\mathrm{min}$ corresponds to the lowest mass a DM halo must have to detach from the expansion of the Universe, to collapse, and to virialize (it is approximately equal to the Jeans mass at recombination).
\end{table}

The CSFR used by \citet{VangioniNitrogen} follows the expression originally derived by \citet{sh2003}, but changing the values of the four free parameters to adjust $\dot\rho_{\star}(z)$ to the most recent observational data. In this work, we include the available GRB data at high redshifts since the UV data will naturally suffer from the selection effect; only the brightest objects are observed.

Although our model is semi-analytic, by adding the redshift dependence to the functions  $<\varepsilon_{\star}>$ and $\tau_\mathrm{s}$, it becomes possible to obtain $\dot\rho_{\star} (z)$ with an adequate behavior within the redshift range where the CSFR data exist. In addition, the way we build $<\varepsilon_{\star}>$ and $\tau_\mathrm{s}$ with the constraints that these functions must satisfy at $z = 0$, shows good agreement with the observational data. In particular, the ratio $<\varepsilon_{\star}(z)>/<\varepsilon_{\star}(z=0)>$ provided by our model is in good agreement with that obtained by \citet{Scoville2017} within the redshift range $[0 - 3]$.

\subsection{Chemical evolution scenario}
\label{CE}
The first chemical evolution models were developed for the framework of the Galaxy by \citet{tinsleylarson1978}, \cite{tinsleylarson1980}, and later by \citet{Matteucci2001}. Their simple model of chemical evolution considers a closed-box evolving system with no inflows or outflows. Also, the IMF is constant in time, the chemical composition of the gas is primordial, and the mixing between the chemical products ejected by stars and the ISM is instantaneous.

We can adapt these concepts, which are the basis of the chemical evolution models of the Galaxy, straightforwardly. The main difference is that in the cosmological scenario, the halos continuously incorporate baryons (primordial gas) from the ambient (Universe). This is described by the function $a_\mathrm{b} (t)$.

Once inside the halos, the gas is removed from the system to form stars at time $t-\tau_\mathrm{m}$. This is described using the CSFR $\dot\rho_{\star}(t-\tau_\mathrm{m})$. Later, the gas returns to the system, at time $t$, when the stars die. A certain fraction of the gas used for the star formation in $t-\tau_\mathrm{m}$ will be retained in the remnant population $m_\mathrm{r}$ that forms in the time $t$. A new generation of stars will be formed at the instant $t$, removing gas from the system, and this processes is repeated in a cycle of continuous gas capture and chemical enrichment of the environment.

To determine the chemical enrichment of a given $i$-element, in addition to the functions $\dot\rho_{\star}$ and $a_\mathrm{b}$, we need to know how much mass of the $i$-element is returned when the star of mass $m$ dies. This is described by the parameter $P_{Z_\mathrm{i\,m}}$ that provides the ``stellar yield'' of the $i$-element.

Once all of these functions and parameters are characterized, we can write a differential equation for the mass density of the $i$-element as:

\begin{multline}\label{chem2}
 \frac{d\rho_\mathrm{g \,i}}{dt} =  \int_{m(t)}^{m_\mathrm{s}}\,[(m - m_\mathrm{r})\,Z_\mathrm{i}\,(t - \tau_\mathrm{m}) + P_{Z_{\mathrm{i\,m}}}]\, \dot\rho_{\star} (t - \tau_\mathrm{m})\\ \varphi (m)\,dm 
 - Z_\mathrm{i}\,\dot\rho_{\star} (t),
\end{multline}
where the term $(m - m_\mathrm{r})\,Z_\mathrm{i}(t - \tau_\mathrm{m})$ accounts for the amount of $i$-element incorporated when the star was born, and which later returns to the ISM (we see that $m_\mathrm{r}\,Z_\mathrm{i}\,(t - \tau_\mathrm{m})$ is the part of the $i$-element retained into the remnant). We note that the resulting $\rho_\mathrm{g}$ is dependent on $a_\mathrm{b}(t)$ through Eq. (\ref{gas_total}), and thus the term $Z_\mathrm{i} = \rho_\mathrm{g\,i}/\rho_\mathrm{g}$ takes into account $a_\mathrm{b}(t)$ through $\rho_\mathrm{g}$. The $P_{Z_\mathrm{i\,m}}$ parameter is the mass produced of the $i$-element by a star of mass $m$. The term $Z_\mathrm{i}\,\dot\rho_{\star}(t)$ takes into account the removal of part of the $i$-element to form a new star generation. 

Through the time integration of Eq. (\ref{chem2}), we obtain the mass density $\rho_\mathrm{g\, i}$ of the $i$-element present in the gas contained within the halos. This allows us to determine quantities such as $[\mathrm{X}_\mathrm{i} / \mathrm{H}]$ as a function of redshift (or time) and to compare the results of our model with different observational data. It should be noted that Eq. (\ref{chem2}) incorporates all the physics and constraints discussed in the previous sections.

In order to incorporate the contributions of particular stars, depending on their masses and metallicities, we selected stellar yields. These chemical yields are used to determine the elements that were ejected into the ISM at a given time by a star of a given mass and metallicity. They are calculated through detailed nucleosynthesis computational simulations, considering the main reactions that happen inside the stars. We consider the first stars to be zero-metallicity stars (Pop III); the subsequent more enriched Pop II stars are chosen within a range of different masses and metallicities. Tables \ref{tab01} and \ref{tab02} describe the stellar mass and metallicity ranges from where the chemical yields were chosen. 

\begin{table}[!ht]
\begin{center}
\centering
\caption{Masses selected for Pop III chemical yields}
\label{tab01}
\begin{tabular}{lcll}
\hline
\hline
Model                          & \multicolumn{1}{l}{CL08} & HW10   & HW02 \\ \hline
                          \hline
Metallicity ($Z_{\sun}$) & \multicolumn{3}{c}{Mass ($\mathrm{M}_{\sun}$)}                                                                                  \\ \hline
\multicolumn{1}{c}{$0$}   & 0.85 - 3.0                                              & \multicolumn{1}{c}{10 - 100} & \multicolumn{1}{c}{140 - 260}  \\ \hline
\end{tabular}
\end{center}
{\small 
{\bf Note.} CL08: \citet{campbelllattanzio2008}, HW10: \citet{hegerwoosley2010}, and  HW02: \citet{heger2002nucleosynthetic}.
}
\end{table}

\begin{table}[!ht]
\begin{center}
\centering
\caption{
Masses and metallicities selected for Pop II chemical yields.}
\label{tab02}
\begin{tabular}{ccccc}
\hline
\hline
\multicolumn{1}{l}{Model}                          & K10   & D13       & D14   & CL04     \\ \hline \hline
\multicolumn{1}{l}{Metallicity ($Z_{\sun}$)} & \multicolumn{4}{c}{Mass ($\mathrm{M}_{\sun}$)}  \\ \hline
$10^{-6}$                                     & -     & -         & -         & 13 - 35  \\ \hline
$10^{-4}$                                     & 1 - 6 & -         & 6.5 - 9.0 & 13 - 35  \\ \hline
$10^{-3}$                                     & -     & -         & 6.5 - 9.0 & 13 - 35  \\ \hline
$4 \times 10^{-3}$                            & 1 - 6 & 6.5 - 9.0 & -         & -        \\ \hline
$6 \times 10^{-3}$                            & -     & -         & -         & 13 - 35        \\ \hline
$8 \times 10^{-3}$                            & 1 - 6 & 6.5 - 9.0 & -         & -        \\ \hline
$2 \times 10^{-2}$                            & 1 - 6 & 6.5 - 9.0 & -         & 13 - 35        \\ \hline
\end{tabular}
\end{center}
{\small 
{\bf Note.} K10: \citet{karakas2010}, D13: \citet{doherty2013}, D14: \citet{doherty2014}, and CL04: \citet{chieffilimongi}.
}
\end{table}

Properly modeling chemical yields for Pop III stars is a challenging and complex task. For the range where they become PISNe, we chose to work with results from \citet{heger2002nucleosynthetic}, which are compatible with recent chemical yields calculated by \citet[][hereafter TK18]{Takahashi}, which take into account rotating progenitors. The two models, HW02 and TK18, show no significant differences in the explosive yields for the elements chosen here, except for the large production of N in the TK18 nonmagnetic rotating models. The N behavior can be better understood in a detailed, recently developed model for the cosmological evolution of this element \citep{VangioniNitrogen}. 

For Pop II stars, samples were chosen according to the best combination of mass and metallicity ranges, and also according to the stellar evolution models and parameters used to produce each sample; \citet{karakas2010} and \citet{doherty2013,doherty2014} used the \textrm{MONSTAR} code for stellar evolution \citep{Monstar}, \textrm{OPAL} opacities \citep{OPAL}, and compatible mass-loss models \citep{Reimers1975,VW1993,Bloecker1995}.

The chemical evolution to be presented in the following sections makes use of the following definition:

\begin{equation}
    [\mathrm{X_{i}/H}] = \log_{10}[\mathrm{N(X_{i})/N(H)}]_\mathrm{gas} - \log_{10}[\mathrm{N(X_{i})/N(H)}]_{\sun},
\end{equation}
where $\mathrm{N(X_{i})}$ and $\mathrm{N(H)}$ are respectively the densities for the element $\mathrm{X_{i}}$ and for hydrogen.

\subsubsection{Numerical technique to determine the elements ${\rho_\mathrm{g\,i}}$ and $\rho_\mathrm{g}$}

The solution for the set of differential equations for $\rho_\mathrm{g}$ and $\rho_\mathrm{g\,i}$ is obtained by a sixth-order Runge-Kutta algorithm. As an initial condition, we have zero values for the densities of gas and metals. To integrate the Pop III and Pop II chemical yields, we use a cubic spline interpolation algorithm for each dataset in Tables \ref{tab01} and \ref{tab02}. For example, for the HW10 model (Pop III in Table \ref{tab01}) with a mass range of $10-100\, \mathrm{M}_{\sun}$, we have 18 stellar mass values, nominally, 10, 12, 15, 17, 20, 22, 25, 27, 30, 33, 35, 40, 45, 50, 60, 75, 85, and 100 solar masses. Thus, the determination of the yields of a star with, for example, $70\,\mathrm{M}_{\sun}$, occurs through the interpolation algorithm.

We see that there are no yields in the intervals $]3.0,10[$ and $]100,140[$ (Table \ref{tab01}). Thus, the interpolation algorithm does not include these open sets in the calculation. A similar implementation applies to Pop II yields (Table \ref{tab02}).

The total density of the gas depends on the term $a_\mathrm{b}(t)$, which corresponds to the infall of primordial gas, basically hydrogen and helium. To check the consistency of our results, we determined the maximum numerical discrepancy $(D)$ between $\rho_\mathrm{g}$ and the sum over all chemical species (hydrogen, helium, and metals), which is:

\begin{equation}
    D = \frac{\rho_\mathrm{g}-\rho_\mathrm{H} - \rho_\mathrm{He}-\sum\,\rho_\mathrm{g\,i}}{\rho_\mathrm{g}}.
\end{equation}

The maximum numerical discrepancy in the various time steps is $\left|D\right| = 4.4\times 10^{-7}$.

\subsection{Constructing the models}
\label{methods}

We start the calculation with all the elements for the cosmological scenario and chemical evolution set, and explore two scenarios. The first (model A) considers the chemical evolution of the Universe starting with the Pop III (masses and yields as shown in Table \ref{tab01}) stars. Once the metallicity of the Universe reaches $Z=10^{-6} Z_{\sun}$, no more Pop III stars can be formed. As a consequence, the Pop II stellar branch with $Z=10^{-6} Z_{\sun}$ is born and evolves (masses and yields as shown in Table \ref{tab02}). This second step finishes when the metallicity of the Universe reaches $Z=10^{-4} Z_{\sun}$ and, as a consequence, no more Pop II stars can be formed within the branch $Z=10^{-6} Z_{\sun}$. This process repeats every time the metallicity of the Universe crosses the limits indicated in Tables \ref{tab01} and \ref{tab02}. It should be noted that the stars within the low-metallicity branches cannot form anymore as a consequence of the increase in the chemical enrichment of the Universe. However, stars with low mass born within the lowest-metal branch can still be alive now. Thus, they can coexist with stars of much higher metallicity during part of their lives. 

For the second scenario (model B), we consider the chemical evolution only with Pop II stars. In this case, the first-star generation ($Z=0$) of the Universe was composed of stars with masses and chemical yields similar to the Pop II stars of the branch $Z = 10^{-6} Z_{\sun}$  studied by \citet{chieffilimongi}. The following steps are similar to those described for model A. Both scenarios are generated with the CSFR described in Sect. \ref{subsec:csfr}.

An important observation is that models A and B have different normalizations for the IMF. For model A, the normalization is obtained in the range $0.1-260\, \mathrm{M}_{\sun}$, while model B (Pop II only) is normalized in the interval $0.1-120\, \mathrm{M}_{\sun}$. This choice is due to the nonexistence of the stellar branch $140-260\, \mathrm{M}_{\sun}$ in the Pop II models discussed in the literature.

We describe the evolution of models A and B in Sect. \ref{2.4.1} and Sect. \ref{2.4.2}.

\subsubsection{Model A}
\label{2.4.1}

Model A runs the following steps for the entire calculation. First, the total metallicity of the Universe ($Z_\mathrm{total}$) is used as a guide for the chemical yields of the different classes (or branches) presented in Tables \ref{tab01} and \ref{tab02}. Assuming that the first stars, formed from pristine (H and He only) gas, started to die and enrich the ISM at redshift $z=20$, $Z_\mathrm{total}$ provides values for the evolution of the production of all elements heavier than He for the entire redshift interval. This parameter is then used as a ``switch'' between different metallicities, removing the chemical yields of a given metallicity and successively introducing those of the higher metallicity classes according to Tables \ref{tab01} and \ref{tab02}, and as discussed above.

Abundances of individual chemical elements are then computed for the entire redshift range. $Z_\mathrm{total}$ starts at zero, producing Pop III stars. The higher-mass Pop III stars ($\sim 260\, \mathrm{M}_{\sun}$) start dying first, throwing metals into the ISM and enriching the medium around them. Pop III stars continue to die and enrich the ISM until the medium reaches $10^{-6} Z_{\sun}$. At that point, new Pop III stars stop forming. It is important to note that Pop III stars with masses $\lesssim 0.9 \mathrm{M}_{\sun}$ have longer lifetimes and should be still in their main-sequence phase today, just following the increase in the total metallicity of the Universe, and will participate in subsequent steps of the enrichment of the Universe when they leave the main sequence, along with the contribution of new, higher-metallicity stars formed later than that population, at much lower redshifts.

Once the metallicity of the ISM and/or IGM reaches $10^{-6} Z_{\sun}$, new stars with this metallicity signature start being formed. As they die, the model starts processing the yields from this class of stars, and the same process repeats. The metallicity from the medium increases as the higher-mass stars die first, while lower-mass stars live longer. Even with stars with a higher metallicity ($10^{-4}\, Z_{\sun}$, for example) starting to form, the lower-mass ones will continue their lives unaffected by the external increase in metallicity.

The process continues as the Universe progresses toward the present metallicity, as described in Tables \ref{tab01} and \ref{tab02}. It is important to emphasize that each metallicity class has its own stellar population clock triggered when the Universe's metallicity crosses the various thresholds indicated in Tables \ref{tab01} and \ref{tab02}. Thus, at certain intervals of time, the chemical enrichment of the Universe takes place by the joint action of stars of different metallicity classes.

\subsubsection{Model B}\label{2.4.2}

Model B uses only Pop II yields. The chemical enrichment of the Universe starts from $Z_\mathrm{total} = 0$ at redshift $20$. For the second model, we consider that the stellar branch $Z=10^{-6}\, Z_{\sun}$ also represents the metal-free stars. Thus, chemical enrichment will occur through this class (or branch) until reaching $Z=10^{-4}$, when then the next metallicity class (with $Z=10^{-3}\,Z_{\sun}$) will assume the control of chemical enrichment. The following branches will take the enrichment command at the points indicated in Table \ref{tab02}.

Chemical elements analyzed in this work were selected considering the availability of chemical yields in the literature and observational data available for chemical abundances in damped Lyman-$\alpha$ systems (DLAs) and globular clusters (GCs), as described below (Sect. \ref{results}).

\section{Results and discussion}
\label{results}

\begin{figure*}[!ht]
\centering
\includegraphics[width=\textwidth]{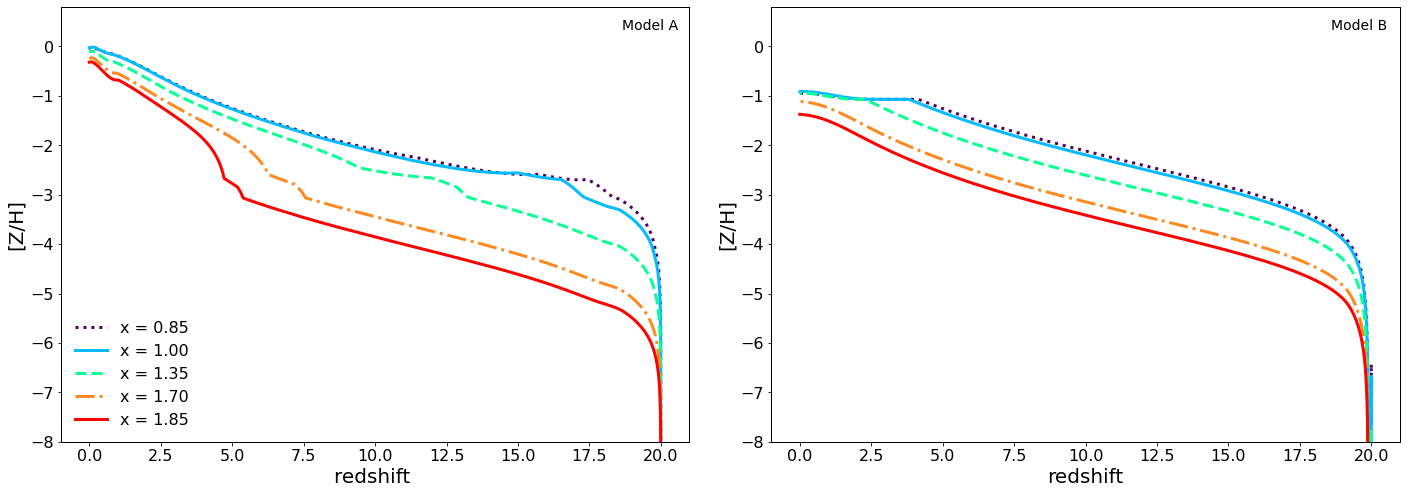}
\caption{Evolution of metallicity for different values of $x$. Left: Model A. Right: Model B. Colors represent different values of $x$ for the IMF.}
\label{fig:ModelA_B}
\end{figure*}

The results from models A and B for $[Z_{\mathrm{total}}/\mathrm{H}]$ are presented in Fig. \ref{fig:ModelA_B}. The results are obtained from the CMW-CSFR with five different IMF values and integrated for all ranges from Pop III to Pop II stars, for model A, and for all ranges of Pop II stars, for model B. We take into account the progressive enrichment of the Universe, and consequently the transition between Pop III stars and the next, more metal-rich Pop II generations until $Z = 0.02\; Z_{\sun}$.

For model A, at redshift $z=20$, the first stars formed from metal-free gas start to die, and the chemical enrichment is very fast. For $x = 0.85$ and $x = 1.00$, the pristine Universe leaves from $Z=0$ to reach $Z=10^{-6}\,Z_{\sun}$ in less than $\sim 4\times 10^{5}\, \mathrm{yr}$, given the higher number of high-mass stars that would form in this scenario. For $x = 1.70$, the same metallicity is reached $\sim 30\,\mathrm{Myr}$ after the death of the first Pop III star, while for $x = 1.85$, it would take $\sim 70\, \mathrm{Myr}$ for the same process to occur. The mean behavior is described by $x = 1.35$, where the Universe would reach $Z=10^{-6}\,Z_{\sun}$ in $\sim 3\,\mathrm{Myr}$. For model B, the same process takes from $\sim 2\, \mathrm{Myr}$ up to $85\, \mathrm{Myr}$, depending on the IMF.

This rapid chemical enrichment in the initial phase can be explained by the metal production of Pop III$-$PISNe, which characterizes a chemical ``flood'' in the high-redshift Universe, in the case of model A. For model B, the chemical enrichment occurs mainly through stars with masses $\sim 30-35\,\mathrm{M}_{\sun}$, and the condition $Z=10^{-6}\,Z_{\sun}$ is reached eight times slower than when considering higher-mass Pop III stars.

Except for N, all other elements are mainly produced by PISNe in the Pop III era. According to \citet{Abia2001}, the metallicity observed at high redshifts can be easily obtained from stellar pregalactic (Pop III) nucleosynthesis by postulating that only $\sim 10^{-2}$ of the total pristine gas is converted into stars. Considering that the star formation efficiency of the CMW scenario is $\sim 0.3$ in the redshift range $[5-20]$, which is $\sim 30$ times larger than the value estimated by \citet{Abia2001}, we verify that adding Pop III stars in the CMW scenario for the CSFR quickly floods the primordial Universe with metals.

Other evidence that PISNe are very efficient in enriching the ISM comes from the work of \citet{MatteucciEarly}, where they show that only 110 to 115 PISNe would be needed to enrich a cubic megaparsec of the IGM to $Z= 10^{-4}\, Z_{\sun}$ (with the index of the IMF varying between 1.35 and 0.5). 

In our model the rate for PISNe can be calculated using the relation:

\begin{equation}
R_\mathrm{PISNe}=\frac{\dot M_{\star}(t)}{<M_\mathrm{PISNe}>}\times \int_{140}^{260} \phi (m)\, m\, dm,
    \label{pisne}
\end{equation}
where $<M_\mathrm{PISNe}>$ is the average mass of the stars that ended their lives as PISNe, and $\dot M_{\star}(t)$ is the CSFR in $\mathrm{M}_{\sun}\,\mathrm{yr}^{-1}$ when $Z = 10^{-6}\, Z_{\sun}$.

If we consider the average mass of PISNe as $\sim 200\,\mathrm{M}_{\sun}$, then $R_\mathrm{PISNe} \sim 6\times 10^{-5}\,\mathrm{yr}^{-1}$ or 1 PISNe every $\sim 16\,000\, \mathrm{yr}$. The number of PISNe needed to enrich the Universe from $Z=0$ to $Z=10^{-6}\,Z_{\sun}$ is $\sim 175$ for our models. 

Similar to \citet{MatteucciEarly}, our work shows the importance of PISNe for rapidly enriching the ISM. \citet{MatteucciEarly} find that $\sim 110-115$ would be needed to enrich a cubic megaparsec of the IGM to $\sim 10^{-4}\,Z_{\sun}$. Our results show that $\sim 175$ PISNe are needed to enrich the medium to $\sim 10^{-6}\,Z_{\sun}$. However, our results involve a volume equivalent to $\sim 10^{5}\,\mathrm{Mpc}^{3}$. Thus, the numerical comparison is not so direct between the two works.

In order to adequately address the question of the rapid contamination of the early Universe, we propose comparing the results with abundances from old GCs. Such GCs (with ages close to the age of the Universe) present an opportunity to explore the chemical and physical conditions of the earliest star-forming environments in the Universe \citep{2011Dotter,FrebelStars}, in other words, they should present a metallicity value very similar to the Universe's mean metallicity at the time they were formed.

\begin{figure*}
\centering
\includegraphics[width=\textwidth]{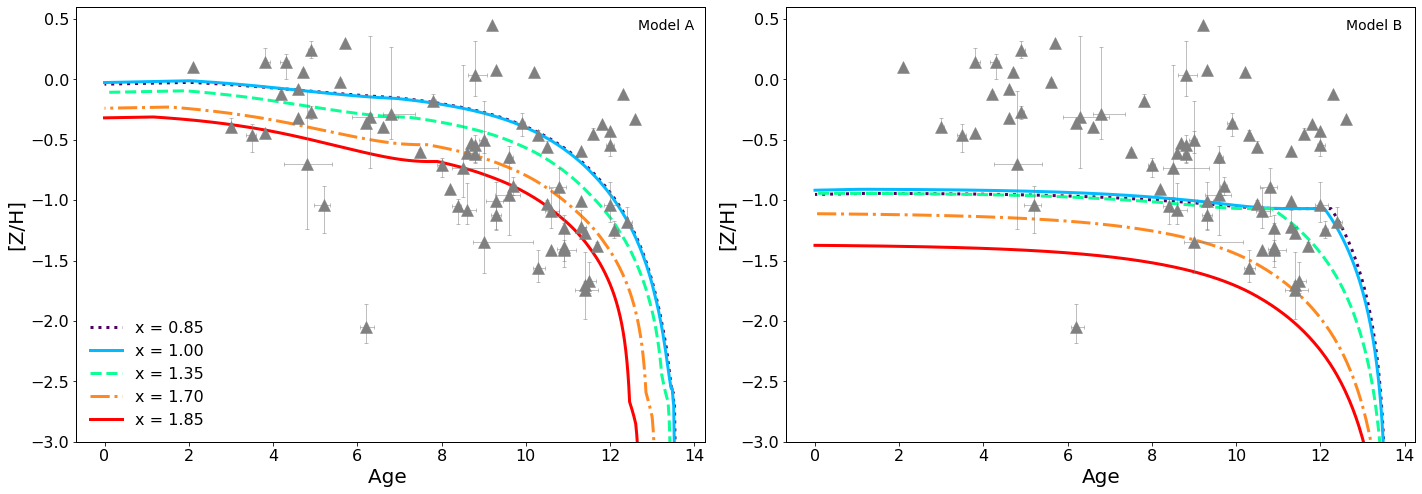}
\caption{Comparison between models A (left) and B (right) with data from GCs from \citet{2007ApJ...660L.117F}, \citet{2013ApJ...765L..12B}, \citet{2002ApJ...572..861C}, \citet{2001Natur.409..691C}, \citet{2011Dotter}, and \citet{2017WK}.}
  \label{fig:GCs_Ztot}
\end{figure*}

Analysis of old GCs' data can provide information about the age and metallicity for the entire cluster, enabling better estimations than for isolated, metal-poor stars, for example. Figure \ref{fig:GCs_Ztot} shows the behavior of $Z_\mathrm{total}$ for the two models compared with observations from GCs. Model A accounts for the majority of observations, regardless of the IMF, while model B is unable to fit the metal abundances for $Z_\mathrm{total}$, even for the lower values of $x$. Although Pop II stars with low metallicity and large masses do indeed play an essential role in the first steps of cosmic enrichment, it is clear that their contribution alone is insufficient to allow the ISM to maintain efficient star formation in order to reach the observed metal abundances along the cosmic history. 

\begin{figure*}[!ht]
\centering
\includegraphics[width=\textwidth]{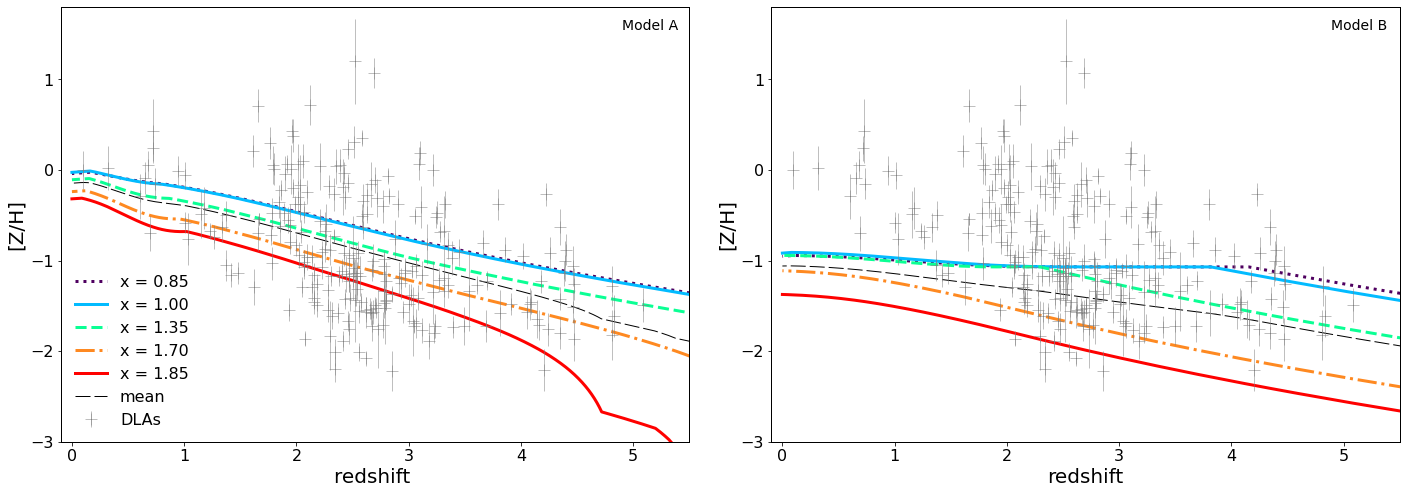}
\caption{Comparison of models A (left) and  B (right) with data from dust-corrected DLAs (gray crosses) from \citet{DeCia2018}. We include an $\alpha$-enhancement correction, with $[Z/\mathrm{H}] = [\mathrm{Fe/H}] + 0.3\,\mathrm{dex}$, as suggested by \citet{2012ApJ...755...89R}.}
  \label{fig:Models_DLAs}
\end{figure*}

Another analysis is performed for the interval $z=[0-6]$, where we compare the results with dust-corrected abundances from DLAs (Fig \ref{fig:Models_DLAs}). The DLAs provide the most accurate measurements of chemical abundances on the gas-phase for the high-redshift Universe \citep{wolfe2005}, and abundances can be determined with errors $\leq 0.1$ dex \citep{Vladilo2002}. The DLAs are also the perfect site for the initial stages of gas cooling and star formation \citep{Maio2015}; they dominate the neutral gas content of the Universe in the redshift interval $z=[0-5]$, and therefore are the most crucial neutral gas reservoir for star formation.

In this redshift interval, regardless of the observations, we expect to see an increase in $[Z_\mathrm{total}/\mathrm{H}]$ with decreasing redshift, with the total metallicity reaching values close to $\sim ~ 0$ (solar) near redshift $z\sim ~ 0$. This behavior is consistent with observations presented in similar contexts \citep{2006A&A...451L..47F,Dave2007,2007MNRAS.376.1465K,10.1111/j.1365-2966.2009.15259.x,VangioniNitrogen}. Also, as pointed out by \citet{calura}, main metal production in spirals and irregulars is always increasing with time.

It is possible to observe that, for model B, for $x < 1.35,$ the abundances reach a maximum at redshift $z\sim 4$  and start decreasing toward $z=0$, while for $x \geq 1.35$, metallicities tend to rise with decreasing redshift. Nevertheless, all the B models remain between $-1.0$ and $-1.5\, \mathrm{dex}$ lower than the expected value for $z \sim 0$.

We also note that Pop II models with $x<1.35$ produce more high-mass stars returning more metals through the CL04 channel. For these IMFs, a bottleneck occurs when the metallicity of the system reaches $8\times 10^{-3}\, Z_{\sun}$ since there is no contribution to chemical enrichment through the CL04 channel. This explains the flat behavior verified in Fig. \ref{fig:GCs_Ztot}.

On the other hand, when taking into account Pop III stars (model A), metallicities increase continuously as redshift approaches $z \sim 0$. For $x = 0.85$, $x = 1.00$, and $x = 1.35$, models reach $[Z_\mathrm{total} / \mathrm{H}]$ close to 0, as expected, while for $x = 1.70$ and $1.85$, the total metallicity is underestimated by approximately $0.25$ to $0.30\, \mathrm{dex}$. 

When comparing results with DLA observations, there are two main problems that are relevant to the interpretation of our results. The first is the high dispersion between points relative to the same (or very close) redshift. There are a variety of models that investigate dispersion in DLAs (see, e.g., \citet{2015MNRAS.452L..36D}), and some authors agree that it happens due to peculiar nucleosynthetic signatures from each system and also due to different star formation histories \citep{Centurion1998,Pettini2000,DZ2002,DeCia2016}, which leads to the production of different amounts of each chemical element. 

According to \citet{DeCia2016}, regardless of the star formation history, the availability of refractory metals in the ISM is a crucial driver of dust production, and DLA galaxies may have a wide range of star formation histories, which in principle are also different from those of the Galaxy \citep{DeCia2016}. Therefore, we plot the mean value between the models with different IMFs, suggesting that results for $x = 0.85$ represent the upper limit (due to the favorable formation of high-mass stars in this model) and $x = 1.85$ as the lower limit (due to the favorable formation of lower-mass stars).

The second problem in comparing results with DLA observations relates to dust depletion. Some chemical elements react with different species, forming molecular compounds that can get trapped on the surface of dust grains and cannot be detected by the observations of abundances in the gas phase, that is to say, abundances would look lower than their actual values. The majority of the results indicate that the behavior of dust depletion on DLAs is complex and varies from system to system
\citep{VladiloSilicon,DeCia2016,DeCia2018}. 

Therefore, we compare our results with dust-corrected DLA metallicities from \citet{DeCia2018}. The author shows that, when including dust corrections, the average DLA metallicities are between $0.4$ and $0.5\, \mathrm{dex}$ higher than without corrections. Where the author provides values for $[\mathrm{Fe/H}]$, we include an $\alpha$-enhancement correction, with $[\mathrm{Z/H}] = [\mathrm{Fe/H}] + 0.3\, \mathrm{dex}$ as suggested by \citet{2012ApJ...755...89R}.

Either way, we reinforce that we aim to demonstrate that Pop III stars are required to represent mean cosmic abundances, which can be straightforwardly observed in Figs \ref{fig:GCs_Ztot} and \ref{fig:Models_DLAs}. The impact of different depletion-corrected methodologies, fitting of the data, and the use of different chemical yields should be addressed in detail in the future.

The knees that appear for model B in Figs. \ref{fig:GCs_Ztot} and \ref{fig:Models_DLAs} are associated with the IMF exponents 0.85 and 1.00. The main reason for that comes from the nonexistence of stars enriching the medium with masses above $9\,\mathrm{M}_{\sun}$ for metallicity classes $4\times 10^{-3}$ and $8\times 10^{-3}$. We see that these $x$ exponents are responsible for forming a higher number of high-mass stars when compared to the other IMFs. This introduces a flat behavior for the $[\mathrm{Z/H}]$ relation for these IMF exponents.

Although there are uncertainties about the mass spectrum of Pop III stars, our results show the importance of this stellar population for reproducing the observational data. There is no good agreement for the metallicity observed for the Universe only with Pop II stars. We see that Pop III stars rapidly raise the metallicity of the Universe between redshifts $\sim 15-20$, mainly due to the HW02 branch. They reinforce chemical enrichment in the range $\sim 5-15$ through the HW10 model and complement the interval $0-5$ through the CL08 channel. The chemical avalanche produced by Pop III stars at high and moderate redshifts acts as a booster so that Pop II stars can add their contribution, through the different branches of Pop II, to the chemical enrichment of the Universe.

\begin{figure*}[!ht]
\centering
\includegraphics[width=\textwidth]{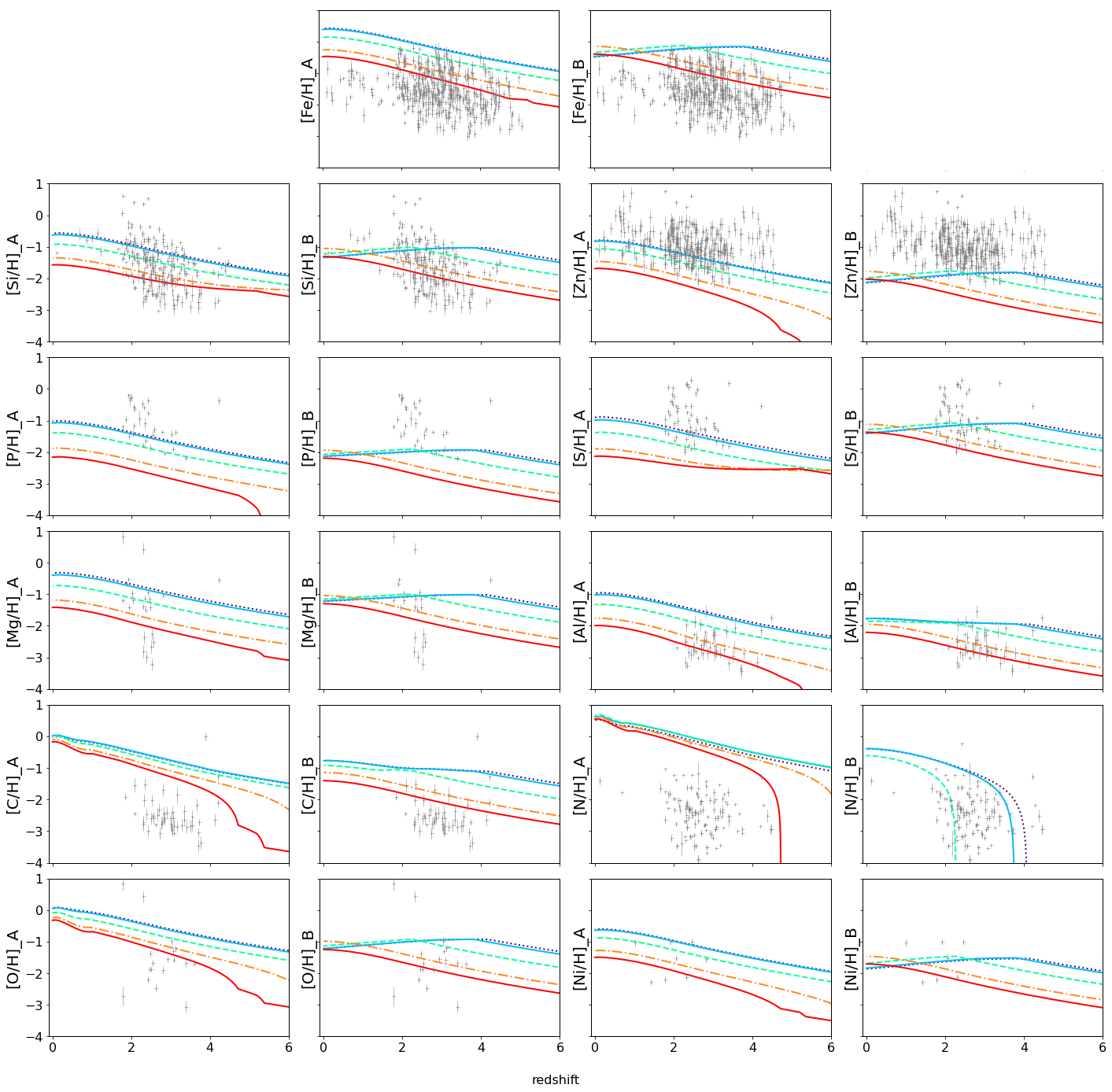}
\caption{Chemical evolution for $\mathrm{Fe,~Si,~Zn, ~Ni,~P, ~Mg, ~Al, ~S, ~C, ~N}$, and $\mathrm{~O}$ since the first stars started to die ($z=20$) until $z=0$. Model A starts with zero-metallicity stars, and as the Universe gets enriched, subsequent Pop II stars with increasing metallicity start to appear, until reaching $Z=2 \times 10^{-2}\, Z_{\sun}$, according to the model described in Sect. \ref{methods}. It is possible to observe the chemical avalanche in the early Universe given by the high production of metals from Pop III stars. As discussed in the text, model B considers only Pop II stars. The gray crosses represent data from \citet{Pettini1997},
\citet{Centurion1998},
\citet{Vladilo1997},
\citet{Pettini2000},
\citet{DZ2002,DZ2003},
\citet{Nissen2004},
\citet{Akerman2005},
\citet{2007ApJ...661...88K},
\citet{Prochaska2007},
\citet{Pettini2008},
\citet{2010ApJ...721....1P},
\citet{Cooke2011},
\citet{Kulkarni2012},
\citet{2012ApJ...755...89R},
\citet{2013MNRAS.435..482J},
\citet{2013ApJ...769...54N},
\citet{2014MNRAS.444..744Z}, and \citet{DeCia2016}. All abundances were rescaled to solar values from \citet{Asplund2009}.}
  \label{fig:Models_grid}
\end{figure*}

\subsection{Properties of yields}

In this section, we present the cosmic chemical evolution for 11 chemical elements for models A and B, compared with data from DLAs taken from the literature (Fig. \ref{fig:Models_grid}), and briefly discuss the main subjects regarding each of the elements. Observational data for other elements are not dust-corrected due to the lack of sufficient data points with enough information for correction (such as $[\mathrm{Zn/Fe}]$ or $[\mathrm{Si/Fe}]$). 

\textit{Iron and silicon:} Fe and Si could be altered by depletion \citep{DeCia2013}. Observational abundances could increase by $\sim 0.5$ dex if depletion is considered in the comparison of the model with DLA data. Details about the methodology used to correct Fe depletion can be checked in the work by \citet{Vladilo2002}. As for Si, \citet{Prochaska2002} show that although it is a refractory element, its depletion is not strong enough to significantly alter the abundances of DLA systems. \citet{VladiloSilicon} show that Si depletion is mild in the ISM and it is expected to be weaker in most DLA systems. The depletion of Fe and Mg are measured for comparison, and it is found that the mean depletion of Si is almost as high as that of Fe, despite Fe being much more depleted than Si in the galactic ISM. They also explain that Si depletion in DLA systems does not correlate with metallicity, unlike Fe, whose depletion rises along with metallicity increase.
    
\textit{Zinc: }
Zn is produced mainly in HNe explosions characterized by a more significant production of Zn, Co, V, and Ti than normal SNe \citep{NomotoHyper}. Stars with $500-1000\,\mathrm{M}_{\sun}$ produce high amounts of Zn compared to O, C, and other metals \citep{Ohkubo2006}. \citet{2006ApJ...653.1145K} suggest that HNe can enhance the production of Zn, and that Zn is considered to be undepleted in DLAs.

\textit{Nickel and phosphorus: }
The lack of Ni observations in DLAs poses a challenge in analyzing this element; nevertheless, SNe Ia produce between $4 \times 10^{-3}$ and $1.4 \times 10^{-2}\, \mathrm{M}_{\sun}$ of Ni, and from $8.5 \times 10^{-5}$ to $4 \times 10^{-4}\, \mathrm{M}_{\sun}$ of P, depending on the specific model \citep{Nomoto1997}, and it is important to observe the outcome of these types of stars in the present model.

\textit{Magnesium: }
Mg is a refractory element, and its depleting effect must be considered. A challenge that arises in Mg determination comes from the saturation of the doublet used for its characterization, leaving only one possible line to provide Mg abundance. Given the problems related to its determination, current observations should be confirmed by additional Mg measurements to conclude if it could have a particular nucleosynthesis effect in DLA systems \citep{DZ2002}.

\textit{Aluminum: }
Al has the strongest metal line transition observed in DLAs, the Al II $\lambda\,1670$ line \citep{Prochaska2002}.\ However, in the majority of systems, the line is heavily saturated and, together with the blending of lines and blending with the Ly-$\alpha$ forest, determining Al abundances can be a real challenge \citep{DZ2003}.

\textit{Sulfur: }
S is considered non-refractory by some authors \citep{Prochaska2002}, but there is still discussion about its actual behavior and if it could be used as a parameter for measuring depletion \citep{Jenkins2009,DeCia2016}.

\textit{Carbon, nitrogen, and oxygen: }
There is an excess in the abundance of C, N, and O. Regarding depletion effects, O is only mildly refractory according to observations of DLAs and is not highly affected by depletion, although it is challenging to observe once the majority of the lines fall into the Ly-$\alpha$ forest and tend to be saturated \citep{Prochaska2002}. On the other hand, C is considered mildly refractory \citep{Prochaska2002}. Once it is a major constituent of interstellar dust \citep{Henry2000,Jenkins2009}, a substantial part of C could be trapped on dust grains. Also, the lack of observations of C in DLAs is a problem \citep{Jenkins2009}. N does not exhibit progressively stronger depletions \citep{Jenkins2009} and appears to be better represented by model B, that is, by the behavior of Pop II stars.

There are, however, other physical processes that participate in the C, N, and O production dynamics, which could be interfering with the results.
\citet{Jenkins2009} shows that, depending on the case, the consumption of O for producing oxides and silicates is not consistent with results for differential depletion for this element. The lack of O in gas-phase observations is much higher than what is needed for producing these silicates and oxides, and it is very hard to correlate the lack of O in the ISM with models of interstellar grain production. The author suggests that the formation of compounds involving elements such as H or C could play an important role in taking these elements from the ISM. Therefore, although cooling processes considerably demand C, N, and O for gas cooling and fragmentation, the grain formation processes do not entirely solve the problem for all three of these overabundant elements.
An interesting result from \citet{Ioppolo} suggests that O is incorporated in the form of amorphous $\mathrm{H_{2}O}$ ice on the grain surfaces. Work recently developed by \citet{Loeb2016-1} suggests that there is a possibility that carbon-enhanced metal-poor (CEMP) stars from the second generation of stars could host or have hosted planetary systems in their habitable zones. The planets would likely have a major C component in their composition. Also, the degree of C enhancement in CEMP stars has been shown to notably increase as a function of decreasing metallicity \citep{LoebRef2}, that is, the C enhancement in this type of star is likely much higher in the primordial Universe.
\citet{LoebRef1} also show that the abundance of water vapor in gas clouds in the Galaxy holds $\sim 0.1\%$ of the available O. 

Lastly, the DLA data have scattering larger than that produced by our models. This result can be associated with several effects. For example, DLA data depend on the characteristics of the host galaxies. As our model is semi-analytic, it cannot resolve individual galaxies. In addition, the CSFR obtained in our models ends up being a weighted average over the halo masses through the formulation presented in Sect. \ref{subsec:csfr}. On the other hand, diffusion is a physical process that can transport metals away from their production sites, which could introduce scattering into our models. Aggregate diffusion is a task to be explored in future work.

\subsection{The effect of feedback on our results}

The feedback effects configure in a complex task, and they are beyond the scope of this work. However, through our modeling, we can make some inferences about the possible impacts on our results. In a recent article, \citet{ Lancaster_2021} presented simulations in which one of the main characteristics of feedback effects is the reduction of star formation efficiency. The authors' simulations involved clouds with characteristic radii of $2.5$ to $20\,\mathrm{pc}$. The efficiency reduction achieved in these simulations ranged from 3\%\ to 47\%  over the no-feedback efficiency.

In our case, the reduction of the star formation efficiency implies a decrease in the $\tau_\mathrm{s}$ scale to keep the CSFR adjusted to the observational data. In this case, some chemical elements such as Zn, P, and S would deviate more from the DLA data, especially with the IMF exponent $x\geq 1.70$. A possible way to solve this problem would be the inclusion of other channels for the production of chemical elements, such as the inclusion of hypernova yields and SNe Ia.

\section{Summary and conclusions}\label{final}

The main goal of this work was to investigate cosmic chemical enrichment through the evolution of chemical elements in the redshift interval $0 \le z \le 20$, as well as the contributions of Pop III and Pop II stars to the cosmic enrichment of the Universe. It was achieved by building a cosmic chemical evolution model that couples a semi-analytic cosmological model, which computes the CSFR, to chemical evolution models for the galactic framework. We computed the evolution of production of $\mathrm{Fe,~Si,~Zn, ~Ni,~P, ~Mg, ~Al, ~S, ~C, ~N}$, and $\mathrm{~O}$ and compared our results with observational data taken from DLAs in the redshift interval $[0-6]$ and with GCs.

Our main results show that we can consistently model the evolution of cosmic abundances in the Universe using a semi-analytic approach. Also, the ``chemical avalanche'' on the primordial Universe, which quickly enriches the medium and provides conditions for Pop II stars to appear, is consistent with the literature on Pop III stars' behavior and chemical evolution models \citep{heger2002nucleosynthetic,hegerwoosley2010,Takahashi}.

Regarding the behavior of Pop III and Pop II stars separately, the main difference appears in the behavior of abundances toward $z=0$. At the same time, our model considering regular intermediate and high-mass Pop II stars (model B) shows a decrease in the abundances (except for N and $Z_\mathrm{total}$), while the model including very massive Pop III stars (model A) reproduces increasing abundances as redshift decreases, which is consistent with observations and similar models in the literature. Model A also offers a better fit of $Z_\mathrm{total}$ to GC data than model B.

We conclude stating that model A, where the inclusion of Pop III stars appears as the main difference, presents a very good description of mean chemical values across the studied redshift range and the key behavior for the evolution of cosmic abundances in the Universe. Our main results are summarized below:

\begin{itemize}
    \item The chemical enrichment process in the early Universe occurs very quickly regardless of the stellar population. The pristine Universe reaches $Z = 10^{-6}\, Z_{\sun}$ in $\sim 3.0\,\mathrm{Myr}$ for the model with both Pop II and Pop III stars and with IMF $1.35$, and $\sim 25\,\mathrm{Myr}$ for the model with only Pop II stars (with $x=1.35$). However, when considering only high-mass Pop II stars, the metals are quickly consumed, and the scenario cannot represent chemical abundances at lower redshifts.

    \item Abundances from GCs for $Z_\mathrm{total}$ are consistently represented by the model with both Pop II and Pop III stars, while the model without Pop III stars is unable to represent observational data, regardless of the IMF. 
    
    \item Abundances from DLAs for $Z_\mathrm{total}$ are consistently represented by our model with Pop III and Pop II stars. When comparing the model with abundances corrected for dust depletion and alpha enhancement, the observations show proper accordance with the model considering both Pop II and Pop III stars, while the model with only Pop II stars cannot account for the behavior of metals toward $z=0$. 
   
    \item Regarding the modeling for other elements, there are a few deviations in the results when comparing the models with data from DLAs. However, the combination of mechanisms needed to improve the results is self-completing and can be easily understood, such as the absence of some mechanisms SNe Ia, HNe, dust depletion affecting observational data, and the combination of yields from Pop II stars. HNe and maybe a higher-mass branch of stars $\sim 500-1000\, \mathrm{M}_{\sun}$, \citep{Ohkubo2006} should improve results for Zn, P, and Ni without raising O and C. In principle, these mechanisms are all consistent with each other and will be studied in a subsequent work.
    
    \item The reason for the overabundances of C, N, and O shown in our results remains an open question. New observations focusing on depletion processes in the ISM could explain the overabundances found in the present work (and/or the lack of these elements in the ISM).
    
\end{itemize}

Altogether, our results indicate that the evolution of chemical abundances in the cosmological framework can be consistently tracked. Our most important result shows that Pop III stars' contribution to the Universe's chemical history should be better understood, and observational campaigns with instruments capable of actually identifying these objects should be seriously considered and implemented. Pop III observations are a long-awaited result, and a firm detection will shed new light on the cosmic history in earlier times. The other questions raised in this paper are being studied and will be the subject of forthcoming works.

\section*{Acknowledgements}
We thank the Brazilian agency FAPESP for support under the thematic project 2014/11156-4. LCC would like to thank the Coordena\c{c}\~ao de Aperfei\c{c}oamento de Pessoal de N\'ivel Superior (CAPES) - Finance code 001 - for a graduate research fellowship. ODM and CAW thank CNPq for partial financial support (grants 303350/2015-6 and 313597/2014-6, respectively).

\bibliographystyle{aa} 
\bibliography{main} 

\end{document}